\documentclass[prl,twocolumn,final]{revtex4}
\usepackage{amsmath}
\usepackage{amsfonts}

\input{tcilatex}
\newtheorem{acknowledgement}[theorem]{Acknowledgement}
\begin{document}

\preprint{}
\title[Effective Simulation of Quantum Entanglement]{Effective Simulation of
Quantum Entanglement Based on A Single-photon Field Modulated with
Pseudorandom Phase Sequences}
\author{Jian Fu, Yi Hu, and Shuo Sun}
\affiliation{State Key Lab of Modern Optical Instrumentation, Department of Optical
Engineering, Zhejiang University, Hangzhou 310027, China}

\begin{abstract}
We demonstrate that a single-photon field modulated with $n$\ different
pseudorandom phase sequences (PPSs) can constitute a $2^{n}$-dimensional
Hilbert space that contains tensor product structure. By using the single
photon field modulated with PPSs, we discuss effective simulation of Bell
states and GHZ state, and apply both correlation analysis and von Neumann
entropy to characterize the simulation. We obtain similar results with the
cases in quantum mechanics and find that the conclusions can be easily
generalized to $n$ quantum particles. The research on simulation of quantum
entanglement may be important, for it not only provides useful insights into
fundamental features of quantum entanglement, but also yields new insights
into quantum computation.
\end{abstract}

\date{today}
\startpage{1}
\email{jianfu@zju.edu.cn}
\maketitle

The phenomenon of quantum entanglement is perhaps the most fascinating and
important feature\textbf{\ }of quantum theory. On the one hand, quantum
entanglement underlies many of the most curious and controversial aspects of
the quantum mechanical description of the world. On the other hand, quantum
entanglement is widely appreciated as the essential ingredient of a quantum
computer \cite{Nielsen,Bennett,Jozsa}. Quantum entanglement is first
introduced by Einstein, Podolsky, and Rosen as most noticeable the EPR
paradox \cite{Einstein}, which is at the origin of quantum nonlocality. Bell
proposed a remarkable inequality imposed by a local hidden variable theory,
which enables an experimental test on the quantum nonlocality \cite{Bell}.

Simulating quantum computation and entanglement using optical setups is
investigated theoretically and experimentally \cite%
{Cerf,Lee1,Spreeuw,Spreeuw2,Matias,Francisco,Massar}. A quantum bit can be
represented by a distinct path or space mode of a single-photon field in an
interferometric setup, which was treated as classical optics analogies in
the most of the cited papers. However, $2^{n}$ basis states of a $n$-qubit
system must be represented by $2^{n}$ distinct paths or modes of a
single-photon field. These simulations are not effective because of the
exponential increase of the required physical resources with the number of
quantum bits \cite{Cerf}. This drawback can be traced back to a lack of a
rigorous tensor-product structure in terms of subsystems \cite{Jozsal,Blume}%
. It is usually considered that the quantum entanglement between two or more
spatially separable particles can hardly be effectively simulated by using
different degrees of freedom of a single-photon field.

In this letter, we explore an effective simulation of several (say $n$)
quantum bits by using multiple beams splitted from a single-photon field 
\cite{Cerf} modulated with pseudorandom phase sequences (PPSs). Based on the
properties of PPSs, we demonstrate that the $n$ beams modulated with $n$\
different PPSs constitute a $2^{n}$-dimensional Hilbert space which contains
a tensor product structure. This is the essential difference between our
proposal and the other classical simulation explorations \cite%
{Cerf,Lee1,Spreeuw,Spreeuw2,Matias,Francisco,Massar}. By using an optical
interferometric setup, PPSs naturally bring on not only random measurement
results, also an ensemble model to define the ensemble average and
correlation functions. Therefore, a single-photon fock state and random
photon statistics are totally unnecessary in our proposal. We first discuss
the effective simulation of Bell states and GHZ state, then apply both the
correlation analysis and von Neumann entropy to characterize them. We obtain
similar results with the cases in quantum mechanics and find that these
conclusions can be easily generalized to $n$ quantum particles.

The PPSs in our proposal derive from orthogonal pseudorandom sequences,
which have been widely applied to Code Division Multiple Access (CDMA)
communication technology as a way to distinguish different users \cite%
{CDMA,PS}. A set of pseudorandom sequences is generated by using a shift
register guided by a Galois field GF($p$), that satisfies orthogonal,
closure and balance properties \cite{PS}. In this letter, we consider an
m-sequence of period $N-1$ ($N=p^{s}$) generated by a primitive polynomial
of degree $s$ over GF($p$) and apply it to 4-ary phase shift modulation,
which has been a well-known modulation format in wireless and optical
communications \cite{CDMA}. A scheme is proposed to generate a PPS set $\Xi
=\left\{ \lambda ^{\left( 0\right) },\lambda ^{\left( 1\right) },\ldots
,\lambda ^{\left( N-1\right) }\right\} $ over GF($4$). $\lambda ^{\left(
0\right) }$ is an all-$0$ sequence and other sequences can be generated by
using the method as follows: (1) given a primitive polynomial of degree $s$
over GF($4$) \cite{William}, a base sequence of a length $4^{s}-1$ is
generated by using Linear Feedback Shift Register; (2) other sequences are
obtained by cyclic shifting of the base sequence; (3) by adding zeros to the
sequences,\ the occurrence of any element equals to $4^{s-1}$; (4) mapping
the elements of the sequences to $\left[ 0,2\pi \right] $: $0$ mapping $0$, $%
1$ mapping $\pi /2$, $2$ mapping $\pi $, and $3$ mapping $3\pi /2$. The PPS
in our proposal represents a sort of additional degree of freedom, which not
only allows to render the distinct feature of different beams splitted from
a single-photon field, also provides a remarkably rich tensor product
structure \cite{Fu2}.

We first consider the two orthogonal modes (polarization or transverse) of a
single-photon field, which are denoted by $\left\vert 0\right\rangle $\ and $%
\left\vert 1\right\rangle $, respectively. Thus any qubit state $|\varphi
\rangle =\alpha |0\rangle +\beta |1\rangle $\ can be expressed by mode
superposition of the single-photon field, where $\left\vert \alpha
\right\vert ^{2}+\left\vert \beta \right\vert ^{2}=1$\ $\left( \alpha ,\beta
\in 
\mathbb{C}
\right) $. Obviously, all the mode superposition states could span a Hilbert
space, where we can employ unitary transformations to transform the
simulation states. Further, we discuss a single-photon field with multiple
PPSs to simulate multiparticle quantum system. Chosen any two PPSs of $%
\lambda ^{\left( a\right) }$ and $\lambda ^{\left( b\right) }$ from the set $%
\Xi $, we can obtain two simulation states expressed as following 
\begin{eqnarray}
\left\vert \psi _{a}\right\rangle &\equiv &e^{i\lambda ^{\left( a\right)
}}\left( \alpha _{a}\left\vert 0\right\rangle +\beta _{a}\left\vert
1\right\rangle \right) ,  \notag \\
\left\vert \psi _{b}\right\rangle &\equiv &e^{i\lambda ^{\left( b\right)
}}\left( \alpha _{b}\left\vert 0\right\rangle +\beta _{b}\left\vert
1\right\rangle \right) .  \label{eq1}
\end{eqnarray}%
According to the properties of the PPSs and the Hilbert space, we can define
the inner product of the two states and obtain the orthogonal property in
our simulation, 
\begin{eqnarray}
\left\langle \psi _{a}|\psi _{b}\right\rangle &=&\frac{1}{N}%
\dsum\limits_{k=1}^{N}e^{i\left( \lambda _{k}^{\left( b\right) }-\lambda
_{k}^{\left( a\right) }\right) }\left( \alpha _{a}^{\ast }\alpha _{b}+\beta
_{a}^{\ast }\beta _{b}\right)  \notag \\
&=&\left\{ 
\begin{array}{cc}
1, & a=b, \\ 
0, & a\neq b.%
\end{array}%
\right.  \label{eq2}
\end{eqnarray}%
where $\lambda _{k}^{\left( a\right) },\lambda _{k}^{\left( b\right) }$ are
the $k$th units of $\lambda ^{\left( a\right) },\lambda ^{\left( b\right) }$%
, respectively. The orthogonal property supports to construct the tensor
product structure of the multiple states.

Assume two Hilbert spaces $w$ and $v$ spanned by the states $\left\vert \psi
_{a}\right\rangle $ and $\left\vert \psi _{b}\right\rangle $, any linear
combinations of the elements in the direct product space of $w\otimes v$
remain in the same space. We define the two orthogonal modes modulated with
the PPS $\lambda ^{\left( a\right) }$ as the orthonormal bases of the space
of $w$, expressed as $\left\vert 0_{a}\right\rangle \equiv e^{i\lambda
^{\left( a\right) }}\left\vert 0\right\rangle $ and $\left\vert
1_{b}\right\rangle \equiv e^{i\lambda ^{\left( b\right) }}\left\vert
1\right\rangle $. Using the same notion, the orthonormal bases of the space
of $v$ are expressed as $\left\vert 0_{b}\right\rangle \equiv e^{i\lambda
^{\left( b\right) }}\left\vert 0\right\rangle $ and $\left\vert
1_{b}\right\rangle \equiv e^{i\lambda ^{\left( b\right) }}\left\vert
1\right\rangle $.\ The four orthonormal bases are thus independent and
distinguishable. Then the orthonormal bases of the direct product space of $%
w\otimes v$ can be expressed as $\left\{ \left\vert 0_{a}\right\rangle
\otimes \left\vert 0_{b}\right\rangle ,\left\vert 0_{a}\right\rangle \otimes
\left\vert 1_{b}\right\rangle ,\left\vert 1_{a}\right\rangle \otimes
\left\vert 0_{b}\right\rangle ,\left\vert 1_{a}\right\rangle \otimes
\left\vert 1_{b}\right\rangle \right\} $. Further, we can obtain the
following tensor product properties \cite{Nielsen}: (1) for any scalar $z$,
the elements $\left\vert \psi _{a}\right\rangle ,\left\vert \psi
_{b}\right\rangle $ in the spaces of $w$ and $v$, respectively, satisfy $%
z\left( \left\vert \psi _{a}\right\rangle \otimes \left\vert \psi
_{b}\right\rangle \right) =\left( z\left\vert \psi _{a}\right\rangle \right)
\otimes \left\vert \psi _{b}\right\rangle =\left\vert \psi _{a}\right\rangle
\otimes \left( z\left\vert \psi _{b}\right\rangle \right) $; (2) in the
space of $w\otimes v$, the direct product of the combinations of elements
equals to the combination of the direct products of elements, $\left(
\left\vert \psi _{a}\right\rangle +\left\vert \psi _{a}^{\prime
}\right\rangle \right) \otimes \left( \left\vert \psi _{b}\right\rangle
+\left\vert \psi _{b}^{\prime }\right\rangle \right) =\left\vert \psi
_{a}\right\rangle \otimes \left\vert \psi _{b}\right\rangle +\left\vert \psi
_{a}\right\rangle \otimes \left\vert \psi _{b}^{\prime }\right\rangle
+\left\vert \psi _{a}^{\prime }\right\rangle \otimes \left\vert \psi
_{b}\right\rangle +\left\vert \psi _{a}^{\prime }\right\rangle \otimes
\left\vert \psi _{b}^{\prime }\right\rangle $. Using the same notion, we can
construct a $2^{n}$-dimensional direct product space of $\left\vert \psi
\right\rangle ^{\otimes n}\equiv \left\vert \psi _{1}\right\rangle \otimes
\ldots \otimes \left\vert \psi _{n}\right\rangle $\ by using the states $%
\left\vert \psi _{1}\right\rangle ,\ldots ,\left\vert \psi _{n}\right\rangle 
$ of a single-photon field\ with $n$\ PPSs \cite{Fu2}.

Now we first discuss the correlation analysis of the effective simulation of
quantum entanglement and demonstrate the nonlocal correlation with Bell's
inequality and equality criterion. As a contrast, we first study the product
state of two states expressed as Eq. (\ref{eq1}), which corresponds to no
entanglement in quantum mechanics. In order to perform the correlation
analysis, we need an orthogonal projection measurement. The result of
projection measurement on $\left\vert \psi \right\rangle $ can be obtained 
\begin{eqnarray}
\overline{P}\left( \theta \right)  &=&\left\langle \psi \left( \theta
\right) \right\vert P\left\vert \psi \left( \theta \right) \right\rangle  
\notag \\
&=&\left( 
\begin{array}{cc}
\alpha ^{\ast }e^{i\theta } & \beta ^{\ast }%
\end{array}%
\right) \left( 
\begin{array}{cc}
0 & 1 \\ 
1 & 0%
\end{array}%
\right) \left( 
\begin{array}{c}
\alpha e^{-i\theta } \\ 
\beta 
\end{array}%
\right)   \notag \\
&=&\beta \alpha ^{\ast }e^{i\theta }+\alpha \beta ^{\ast }e^{-i\theta }.
\label{eq3}
\end{eqnarray}%
For convenience, the superposition coefficients $\alpha ,\beta $ are given
by $1/\sqrt{2}$, we obtain $\overline{P}\left( \theta \right) =\cos \left(
\theta \right) $. By using the projection measurement, we propose a
correlation analysis scheme in which the projection measurement of the basis 
$\left\vert \pm \right\rangle =\left\vert 0\right\rangle \pm \left\vert
1\right\rangle $ is performed on each of the fields, as shown in Fig. 1. $%
\overline{P}\left( \theta _{a},k\right) =\cos \theta _{a}$ and $\overline{P}%
\left( \theta _{b},k\right) =\cos \theta _{b}$ are the measurement results
in the $k$th sequence unit for the fields $\left\vert \psi _{a}\right\rangle 
$ and $\left\vert \psi _{b}\right\rangle $, respectively. The PPSs of $%
\lambda ^{\left( a\right) }$ and $\lambda ^{\left( b\right) }$ do not affect
the measurement results because they only contribute to the total phases. We
can easily obtain the correlation function $E\left( \theta _{a},\theta
_{b}\right) =\cos \theta _{a}\cos \theta _{b}$, which is similar to the case
of quantum product state.

Further, we consider that the modes $\left\vert 1\right\rangle $ of the
states $\left\vert \psi _{a}\right\rangle $ and $\left\vert \psi
_{b}\right\rangle $ are exchanged by using a mode exchanger constituted by
mode splitters and combiners (Fig. 2), which can be realized by a
polarization beam splitter (for polarization modes, Fig. 3) or a multimode
directional coupler (for transverse modes) \cite{Fu}. We obtain the states
as following 
\begin{eqnarray}
\left\vert \psi _{a}^{\prime }\right\rangle &=&\frac{1}{\sqrt{2}}e^{i\lambda
^{\left( a\right) }}\left( \left\vert 0\right\rangle +e^{i\gamma ^{\left(
a\right) }}\left\vert 1\right\rangle \right) ,  \notag \\
\left\vert \psi _{b}^{\prime }\right\rangle &=&\frac{1}{\sqrt{2}}e^{i\lambda
^{\left( b\right) }}\left( \left\vert 0\right\rangle +e^{i\gamma ^{\left(
b\right) }}\left\vert 1\right\rangle \right) ,  \label{eq4}
\end{eqnarray}%
where the relative phase sequences (RPSs) $\gamma ^{\left( a\right)
}=-\gamma ^{\left( b\right) }=\lambda ^{\left( b\right) }-\lambda ^{\left(
a\right) }$, and $\gamma ^{\left( a\right) }+\gamma ^{\left( b\right) }=0$.
We obtain the results of the fields in the projection measurement $\overline{%
P}\left( \theta _{a},k\right) =\cos \left( \theta _{a}+\gamma _{k}^{\left(
a\right) }\right) $, $\overline{P}\left( \theta _{b},k\right) =\cos \left(
\theta _{b}+\gamma _{k}^{\left( b\right) }\right) $, where $\gamma
_{k}^{\left( a\right) },\gamma _{k}^{\left( b\right) }$ are the $k$th units
of the RPSs $\gamma ^{\left( a\right) },\gamma ^{\left( b\right) }$,
respectively. Then we obtain the correlation function 
\begin{eqnarray}
E\left( \theta _{a},\theta _{b}\right) &=&\frac{1}{NC}\sum_{k=1}^{N}%
\overline{P}\left( \theta _{a},k\right) \overline{P}\left( \theta
_{b},k\right)  \notag \\
&=&\cos \left( \theta _{a}+\theta _{b}\right) ,  \label{eq5}
\end{eqnarray}%
where $C=1/2$ is the normalized coefficient. The correlation function is
similar to the case of Bell state $\left\vert \Psi ^{+}\right\rangle $.
Therefore we consider the states in Eq. (\ref{eq4}) as the single-photon
field simulation of the Bell state $\left\vert \Psi ^{+}\right\rangle $. We
substitute the correlation functions above into Bell inequality (CHSH
inequality) \cite{CHSH}%
\begin{eqnarray}
\left\vert B\right\vert &=&\left\vert E\left( \theta _{a},\theta _{b}\right)
-E\left( \theta _{a},\theta _{b}^{\prime }\right) +E\left( \theta
_{a}^{\prime },\theta _{b}^{\prime }\right) +E\left( \theta _{a}^{\prime
},\theta _{b}\right) \right\vert  \notag \\
&=&2\sqrt{2}>2,  \label{eq6}
\end{eqnarray}%
where $\theta _{a},\theta _{a}^{\prime },\theta _{b}$ and $\theta
_{b}^{\prime }$ are $\pi /4,-\pi /4,0$ and $\pi /2$, respectively, when
Bell's inequality is maximally violated.

Another Bell state $\left\vert \Psi ^{-}\right\rangle $ differs from $%
\left\vert \Psi ^{+}\right\rangle $ by $\pi $ phase. Similarly, we obtain
the simulation of the Bell state $\left\vert \Psi ^{-}\right\rangle $
expressed as $\left\vert \psi _{a}^{\prime }\right\rangle =e^{i\lambda
^{\left( a\right) }}\left( \left\vert 0\right\rangle +e^{i\gamma ^{\left(
a\right) }}\left\vert 1\right\rangle \right) /\sqrt{2}$, $\left\vert \psi
_{b}^{\prime }\right\rangle =e^{i\lambda ^{\left( b\right) }}\left(
\left\vert 0\right\rangle +e^{i\left( \gamma ^{\left( b\right) }+\pi \right)
}\left\vert 1\right\rangle \right) /\sqrt{2}$. By performing the
transformation $\sigma _{x}:\left\vert 0\right\rangle \leftrightarrow
\left\vert 1\right\rangle $ on $\left\vert \psi _{b}^{\prime }\right\rangle $
of the simulation of $\left\vert \Psi ^{\pm }\right\rangle $, we obtain the
simulation of the Bell state $\left\vert \Phi ^{+}\right\rangle $ expressed
as $\left\vert \psi _{a}^{\prime }\right\rangle =e^{i\lambda ^{\left(
a\right) }}\left( \left\vert 0\right\rangle +e^{i\gamma ^{\left( a\right)
}}\left\vert 1\right\rangle \right) /\sqrt{2}$, $\left\vert \psi
_{b}^{\prime }\right\rangle =e^{i\lambda ^{\left( b\right) }}\left(
e^{i\gamma ^{\left( b\right) }}\left\vert 0\right\rangle +\left\vert
1\right\rangle \right) /\sqrt{2}$, and of $\left\vert \Phi ^{-}\right\rangle 
$ expressed as $\left\vert \psi _{a}^{\prime }\right\rangle =e^{i\lambda
^{\left( a\right) }}\left( \left\vert 0\right\rangle +e^{i\gamma ^{\left(
a\right) }}\left\vert 1\right\rangle \right) /\sqrt{2}$, $\left\vert \psi
_{b}^{\prime }\right\rangle =e^{i\lambda ^{\left( b\right) }}\left(
e^{i\left( \gamma ^{\left( b\right) }+\pi \right) }\left\vert 0\right\rangle
+\left\vert 1\right\rangle \right) /\sqrt{2}$. Then their correlation
functions $E_{\Psi ^{-}}\left( \theta _{a},\theta _{b}\right) =-\cos \left(
\theta _{a}+\theta _{b}\right) $, $E_{\Phi ^{\pm }}\left( \theta _{a},\theta
_{b}\right) =\pm \cos \left( \theta _{a}-\theta _{b}\right) $ are obtained.
To substitute the correlation functions into Eq. (\ref{eq6}), we also obtain
the maximal violation of Bell's inequality. The violation of Bell's
criterion demonstrates the nonlocal correlation of the two beams splitted
from the single-photon field in our simulation, which results from shared
randomness of the PPSs.

The nonlocality of the multipartite entangled GHZ states can in principle be
manifest in a single measurement and need not be statistical as the
violation of Bell inequality that relies on mean values \cite{GHZ}.
Preparing three states $\left\vert \psi _{a}\right\rangle ,\left\vert \psi
_{b}\right\rangle $ and $\left\vert \psi _{c}\right\rangle $ similar to Eq. (%
\ref{eq1}), and circle exchanging the modes $\left\vert 1\right) $ of the
fields, we obtain the states as following%
\begin{eqnarray}
\left\vert \psi _{a}^{\prime }\right\rangle &=&\frac{1}{\sqrt{2}}e^{i\lambda
^{\left( a\right) }}\left( \left\vert 0\right\rangle +e^{i\gamma ^{\left(
a\right) }}\left\vert 1\right\rangle \right) ,  \notag \\
\left\vert \psi _{b}^{\prime }\right\rangle &=&\frac{1}{\sqrt{2}}e^{i\lambda
^{\left( b\right) }}\left( \left\vert 0\right\rangle +e^{i\gamma ^{\left(
b\right) }}\left\vert 1\right\rangle \right) ,  \label{eq7} \\
\left\vert \psi _{c}^{\prime }\right\rangle &=&\frac{1}{\sqrt{2}}e^{i\lambda
^{\left( c\right) }}\left( \left\vert 0\right\rangle +e^{i\gamma ^{\left(
c\right) }}\left\vert 1\right\rangle \right) ,  \notag
\end{eqnarray}%
where the RPSs $\gamma ^{\left( a\right) }=\lambda ^{\left( b\right)
}-\lambda ^{\left( a\right) }$, $\gamma ^{\left( b\right) }=\lambda ^{\left(
c\right) }-\lambda ^{\left( b\right) }$, $\gamma ^{\left( c\right) }=\lambda
^{\left( a\right) }-\lambda ^{\left( c\right) }$, and $\gamma ^{\left(
a\right) }+\gamma ^{\left( b\right) }+\gamma ^{\left( c\right) }=0$. We
obtain the measurement results $\overline{P}\left( \theta _{a},k\right)
=\cos \left( \theta _{a}+\gamma _{k}^{\left( a\right) }\right) $, $\overline{%
P}\left( \theta _{b},k\right) =\cos \left( \theta _{b}+\gamma _{k}^{\left(
b\right) }\right) $, $\overline{P}\left( \theta _{c},k\right) =\cos \left(
\theta _{c}+\gamma _{k}^{\left( c\right) }\right) $ for the states $%
\left\vert \psi _{a}^{\prime }\right\rangle ,\left\vert \psi _{b}^{\prime
}\right\rangle $ and $\left\vert \psi _{c}^{\prime }\right\rangle $ in the
projection measurement, respectively, and the correlation function%
\begin{eqnarray}
E\left( \theta _{a},\theta _{b},\theta _{c}\right) &=&\frac{1}{NC}%
\sum_{k=1}^{N}\overline{P}\left( \theta _{a},k\right) \overline{P}\left(
\theta _{b},k\right) \overline{P}\left( \theta _{c},k\right)  \notag \\
&=&\cos \left( \theta _{a}+\theta _{b}+\theta _{c}\right) ,  \label{eq8}
\end{eqnarray}%
where $C=1/4$ is the normalized coefficient. If $\theta _{a}+\theta
_{b}+\theta _{c}=0$, $E\left( \theta _{a},\theta _{b},\theta _{c}\right) =1$%
. If $\theta _{a}+\theta _{b}+\theta _{c}=\pi $, $E\left( \theta _{a},\theta
_{b},\theta _{c}\right) =-1$. This would manifest the nonlocality in a
single measurement for the simulation of GHZ states. It is noteworthy that
the correlation function sinusoidally oscillates with one of the phases $%
\theta _{a},\theta _{b}$ and $\theta _{c}$, when the other phases fixed.
However, the correlation function equals to zero if based on the correlation
analysis of only two states, because $\gamma ^{\left( a\right) }+\gamma
^{\left( b\right) }+\gamma ^{\left( c\right) }=0$ and the sum of any two
RPSs remains a pseudorandom sequence. These results are completely similar
to the case of GHZ states.

Further, the simulation of GHZ state could be generalized to the case of $n$
particles. Prepared $n$ states similar to Eq. (\ref{eq1}) and circle
exchanged the modes $\left\vert 1\right\rangle $ of the states, the RPSs
satisfy $\gamma ^{\left( 1\right) }+\ldots +\gamma ^{\left( n\right) }=0$.
We obtain the correlation function%
\begin{eqnarray}
E\left( \theta _{1},\ldots ,\theta _{n}\right) &=&\frac{1}{NC}\sum_{k=1}^{N}%
\overline{P}\left( \theta _{1},k\right) \ldots \overline{P}\left( \theta
_{n},k\right)  \notag \\
&=&\cos \left( \theta _{1}+\ldots +\theta _{n}\right) ,  \label{eq9}
\end{eqnarray}%
where $\overline{P}\left( \theta _{i},k\right) =\cos \left( \theta
_{i}+\gamma _{k}^{\left( i\right) }\right) $ is the result of the
single-photon field with $i$th PPSs at the $k$th sequence units in the
projection measurement, and $C=1/2^{n-1}$ is the normalized coefficient.
Using the same notion, the simulation of any other generalized GHZ states
can be obtained and the correlation functions are also similar to the case
of quantum mechanics.

Now we continue to discuss our simulation in another view and try to apply
von Neumann entropy as entanglement measure to characterize the
single-photon field simulation of Bell states and GHZ state. First we define
a simulation state in the direct product space of a single-photon field with
multiple PPSs as following%
\begin{equation}
\left\vert \Psi \right\rangle \equiv \frac{e^{i\lambda ^{\left( s\right) }}}{%
C}\sum_{k=1}^{N}\left( e^{-i\lambda _{k}^{\left( s\right) }}\left\vert \psi
_{1}^{\prime k}\right\rangle \otimes ...\otimes \left\vert \psi _{n}^{\prime
k}\right\rangle \right) ,  \label{eq10}
\end{equation}%
where $C$ is the normalized coefficient, $\lambda ^{\left( s\right)
}=\tsum\nolimits_{i=1}^{n}\lambda ^{\left( i\right) }$ denotes the total
phase sequence with the $k$th unit $\lambda _{k}^{\left( s\right) }$, $%
\left\vert \psi _{i}^{\prime k}\right\rangle $ denotes the mode
superposition of the single-photon field with $i$th PPSs at the $k$th
sequence unit. Then, we introduce the density matrix formulation by using $%
\rho =\left\vert \Psi \right\rangle \left\langle \Psi \right\vert $ and Eq. (%
\ref{eq10}), as following 
\begin{eqnarray}
\rho  &=&\frac{1}{\left\vert C\right\vert ^{2}}\sum_{k=1}^{N}\left(
e^{-i\lambda _{k}^{\left( s\right) }}\left\vert \psi _{1}^{\prime
k}\right\rangle \otimes \ldots \otimes \left\vert \psi _{n}^{\prime
k}\right\rangle \right)   \notag \\
&&\times \sum_{k=1}^{N}\left( e^{i\lambda _{k}^{\left( s\right)
}}\left\langle \psi _{1}^{\prime k}\right\vert \otimes \ldots \otimes
\left\langle \psi _{n}^{\prime k}\right\vert \right) .  \label{eq11}
\end{eqnarray}%
The entanglement of a partly-entangled pure state can be naturally
parameterized by its von Neumann entropy of entanglement \cite{Bennett1}.
Given a pure state $\rho _{ab}$ of two subsystems $a$ and $b$, we define the
reduced density matrices $\rho _{a}=tr_{b}\left( \rho _{ab}\right) $ and $%
\rho _{b}=tr_{a}\left( \rho _{ab}\right) $ for the states of $a$ and $b$,
where the partial trace has been taken over one subsystem, either $a$ or $b$%
. Then the von Neumann entropy of the reduced density matrices is given by $%
S=-tr\left( \rho _{a}\log _{2}\rho _{a}\right) =-tr\left( \rho _{b}\log
_{2}\rho _{b}\right) $. The quantity $S$ ranges from zero for a product
state to $1$ for a maximally entangled pair of two-state particles, and $%
0<S<1$ for a partly entangled pair.

We first discuss the entanglement measure of the product state of two states
expressed as Eq. (\ref{eq1}). We can easily obtain the density matrix $\rho
_{ab}=\left\vert \psi _{a}\right\rangle \left\langle \psi _{a}\right\vert
\otimes \left\vert \psi _{b}\right\rangle \left\langle \psi _{b}\right\vert $%
, and the von Neumann entropy $S=-tr\left( \rho _{a}\log _{2}\left( \rho
_{a}\right) \right) =$\ $-tr\left( \rho _{b}\log _{2}\left( \rho _{b}\right)
\right) =0$. This indicates that no entanglement is involved between the two
states. We then discuss the simulation of Bell state $\left\vert \Psi
^{+}\right\rangle $ expressed as Eq.(\ref{eq4}). We can reduce the
expression as $\left\vert \Psi \right\rangle =e^{i\lambda ^{\left( s\right)
}}\left( \left\vert 00\right\rangle +\left\vert 11\right\rangle \right) /%
\sqrt{2}$, where $\left\vert q_{a}q_{b}\right\rangle ,\left(
q_{a,b}=0,1\right) $\ are the orthonormal bases of the direct product space.
It is worthwhile that the balance property $\tsum\nolimits_{k=1}^{N}e^{i%
\gamma _{k}^{\left( i\right) }}=0$\ of RPSs $\gamma ^{\left( i\right) }$\
results in many items of the direct product of $\left\vert \psi _{i}^{\prime
}\right\rangle $\ disappeared. Then we obtain the reduced density matrix $%
\rho _{ab}=\left( \left\vert 00\right\rangle \left\langle 00\right\vert
+\left\vert 11\right\rangle \left\langle 11\right\vert +\left\vert
00\right\rangle \left\langle 11\right\vert +\left\vert 11\right\rangle
\left\langle 00\right\vert \right) /2$ and $S=-tr\left( \rho _{a}\log
_{2}\left( \rho _{a}\right) \right) =$\ $-tr\left( \rho _{b}\log _{2}\left(
\rho _{b}\right) \right) =1$. It means the two fields of simulation of $%
\left\vert \Psi ^{+}\right\rangle $\ are completely entangled.

Using the same notion, we can obtain similar results for the simulation of
other Bell states and GHZ state. Since the von Neumann entropy gives the
same results for the simulation and quantum entangled states, we prove the
validity of the simulation in a more rigorous way. It should be pointed out
that the phase pseudorandomness provided by PPSs is different from the case
of quantum mixed states. Quantum mixed states result from decoherence and
all coherent superposition items disappeared. Different from the
decoherence, some coherent superposition items remain in the simulation
state due to the constraints of the RPSs, such as $\gamma ^{\left( a\right)
}+\gamma ^{\left( b\right) }=0,\gamma ^{\left( a\right) }+\gamma ^{\left(
b\right) }+\gamma ^{\left( c\right) }=0$ for the simulation of Bell states
and GHZ state, respectively. These remaining items makes the simulation of
quantum entangled pure states possible.

In this letter, we utilize the properties of PPSs to distinguish the beams
splitted from a single-photon field that are even overlapped in same space
and time. A $2^{n}$-dimensional Hilbert space which contains tensor product
structure is spanned by the single-photon field modulated with $n$ PPSs. In
our proposal, the resources required are the PPSs instead of optical/space
modes. An optical beam with a PPS can simulate one quantum particle. It
means that the amount of PPSs grows linearly with the number of quantum
particles. According to the m-sequence theory, the number of PPSs in the set 
$\Xi $ equals to the length of sequences, which means that the time resource
(the length of sequence) required also grows linearly with the number of the
particles. Based on the analysis above, we conclude that one can efficiently
simulate quantum entanglement with linearly growing resources by using our
proposal.

A constructive method for simulating quantum entanglement is presented in
this letter, which has similar mathematical expressions and physical
meanings with the cases in quantum mechanics. In the framework of quantum
mechanics, the overall phase of wavefunction can be ignored for no
contribution to the probability distribution. However, quantum entanglement
must be related to two or more spatially separable quantum particles. We
argue that each quantum particle might be characterized by the wave function
with a unique random phase sequence, namely the quantum particle might own
some unknown intrinsic phase mechanism. The intrinsic phase mechanism could
be proven in atomic Bose-Einstein condensate experiments. If this hypothesis
was established, the nonlocality of quantum entanglement described as a
\textquotedblleft spooky action\textquotedblright\ might be easy to
understand.

In summary, we have demonstrated a new proposal to simulate quantum
entanglement and apply both the correlation analysis and von Neumann entropy
to characterize the simulation. We conclude that quantum entanglement can be
efficiently simulated by using a single-photon field modulated with PPSs.
The research on this simulation may be important, for it not only provides
useful insights into fundamental features of quantum entanglement, but also
yields new insights into quantum computation.

\textbf{Appendix: Some discussions on quantum mechanics}

1. Integrity problems of quantum particles

There is no doubt that a quantum particle is integrity and inseparable.
Similarly, PPSs are also integrity and inseparable. Therefore the
single-photon field with a PPS can simulate the integrity of the quantum
particle. By performing quadrature demodulation scheme, we could obtain the
mode status matrix of the simulating classical fields, based on which we
propose a sequence permutation mechanism to reconstruct the simulated
quantum states \cite{Fu2}. This could simulate the click of a single
particle in Bell-CHSH measurement.

2. The ensemble model

In quantum mechanics, the probability distribution of a quantum particle
cannot be achieved by a single measurement unless a large number of repeated
measurements with the help of the ensemble model and ergodicity hypothesis.
In fact, the simulation can also be regarded as an ensemble model. In the
quadrature demodulation, each code obtained in each sequence unit of a PPS
can be regarded as a single measurement. The sequence number of the PPS's
unit can be used as parameters of the ensemble. Different from the
ergodicity hypothesis of quantum mechanics, the ergodicity of PPS is
determined and much more efficient.

3. Nonlocality problem

In our proposal, the randomness is carried by the optical field in the phase
sequence. This randomness followed with optical fields propagates to
measuring positions, causing measure of randomness. Therefore, there are no
problems of nonlocal in the simulation.

4. Origin of randomness

The most noteworthy are the random origin of quantum mechanics that is
related the probabilities of quantum measurable properties and the
completeness of quantum mechanics. If the randomness of a quantum particle
comes from the local measurement process, it inevitably brings the nonlocal
difficulty.

To avoid this difficulty, we propose that each quantum particle might be
characterized by the wave function with a unique random phase sequence,
namely the quantum particle might own some unknown intrinsic phase
mechanism. The intrinsic phase mechanism might be related to superstring
theory \cite{Green}, which might be the geometric phases acquired by the
particle strings winding around the extra dimensions of space.

In superstring theory, Kaluza-Klein compactification of the extra dimensions
has one important difference from the particle theory version. A closed
string can get wound several times around a rolled up dimension (see Fig.
4). When a string does this, the string oscillations have a winding mode.
The winding modes add a symmetry to the theory not present in particle
physics. A theory with a rolled up dimension with size $R$ turns out to be
equivalent to a theory with a rolled up dimension of size $(L_{s}^{2})/R$
with the winding modes and momentum modes in the extra dimension exchanged ($%
L_{s}$ is the string length scale). 

Now we consider the geometric phase due to the winding modes. The geometric
phase is very similar to the Aharonov-Bohm phase of a charged particle
traversing a loop including a magnetic flux \cite{Shapere}, which is a
nontrivial topological effect in a multiply connected space (see Fig. 5). We
can express the geometric phase abstractly%
\begin{equation}
\beta \left( \tau \right) =i\doint \left\langle \phi \left( \sigma \right)
\right\vert d\left\vert \phi \left( \sigma \right) \right\rangle
=i\int_{0}^{\sigma \left( \tau \right) =2\pi }\left\langle \phi \left(
\sigma \right) \right\vert \frac{d}{d\sigma }\left\vert \phi \left( \sigma
\right) \right\rangle d\sigma 
\end{equation}%
where $\sigma ,\tau $ are the space-/time-like parameters of world volume.
The geometric phase depends only on the winding number of the string around
the rolled up dimension, and independs on the size of the dimension even
rolled up in a circle of radius $(L_{s}^{2})/R$. Considering the string
periodic oscillations, it must be a periodic function of time. When the
geometric effects of the multiple extra dimensions superpose, we can obtain
the overall geometric phase $\beta \left( \tau \right) =\sum_{k=1}^{N}\beta
_{k}(\tau _{k})$, where $N$ is related to the number of extra-dimension
spaces. Through the qualitative analysis of the superposed geometric phase,
we propose three theoretical predictions:

\textbf{Autonomy:} A quantum particle autonomously owns the geometric phase
that randomly varies with time even in the absence of any external
disturbances.

\textbf{Periodic: }The random phase of a quantum particle must be periodic
if the external disturbances are absent.

\textbf{Superposition: }The random phase of a quantum particle must be
superposition of multiple periodic functions, and the maximum count of the
periodic functions should be the number of topological nontrivial
extra-dimension spaces.

\begin{acknowledgement}
Supported by the National Natural Science Foundation of China under Grant No
60407003.
\end{acknowledgement}

\begin{description}
\item[Fig. 1] The scheme of the projection measurement, where the black
blocks denote the projection measurement of the basis $\left\vert \pm
\right\rangle =\left\vert 0\right\rangle \pm \left\vert 1\right\rangle $.

\item[Fig. 2] The mode exchange scheme for two classcial fields, where the
gray blocks denote the mode splitters/combiners.

\item[Fig. 3] An implementation of the scheme for polarization modes
exchange.

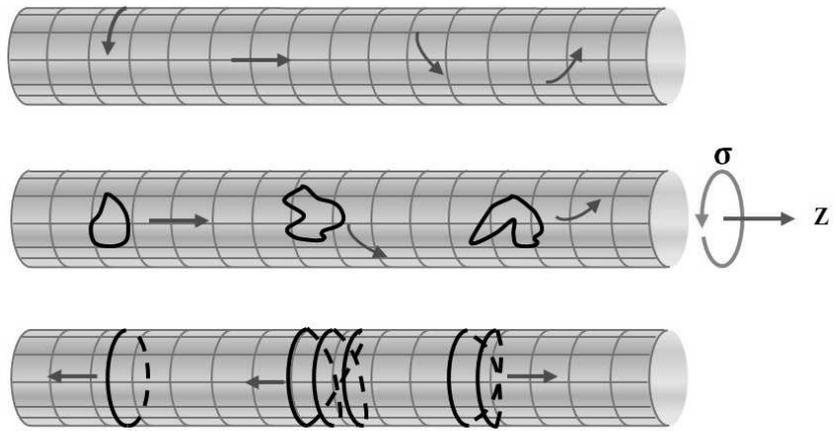
\item[Fig. 4] A closed string can get wound several times around a rolled up
dimension.

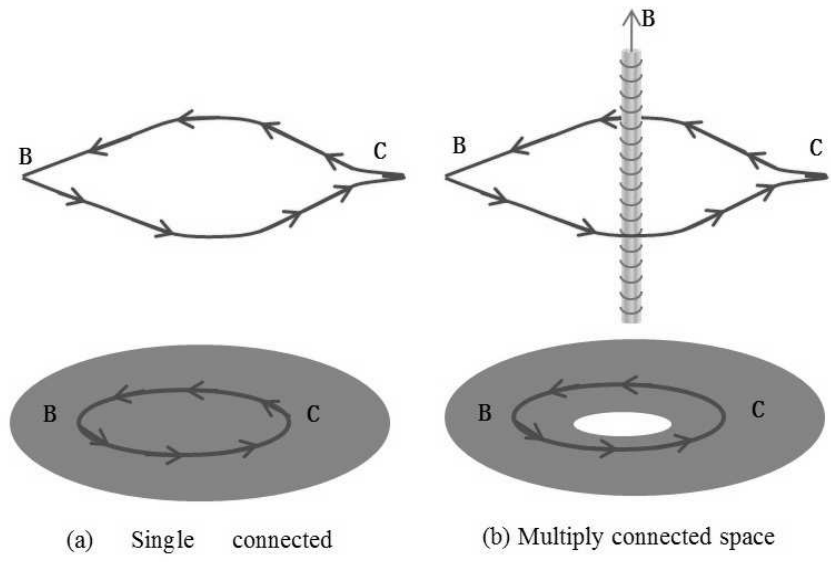
\item[Fig. 5] Aharonov-Bohm phase of a charged particle traversing a loop
including a magnetic flux, which is a nontrivial topological effect in a
multiply connected space.
\end{description}

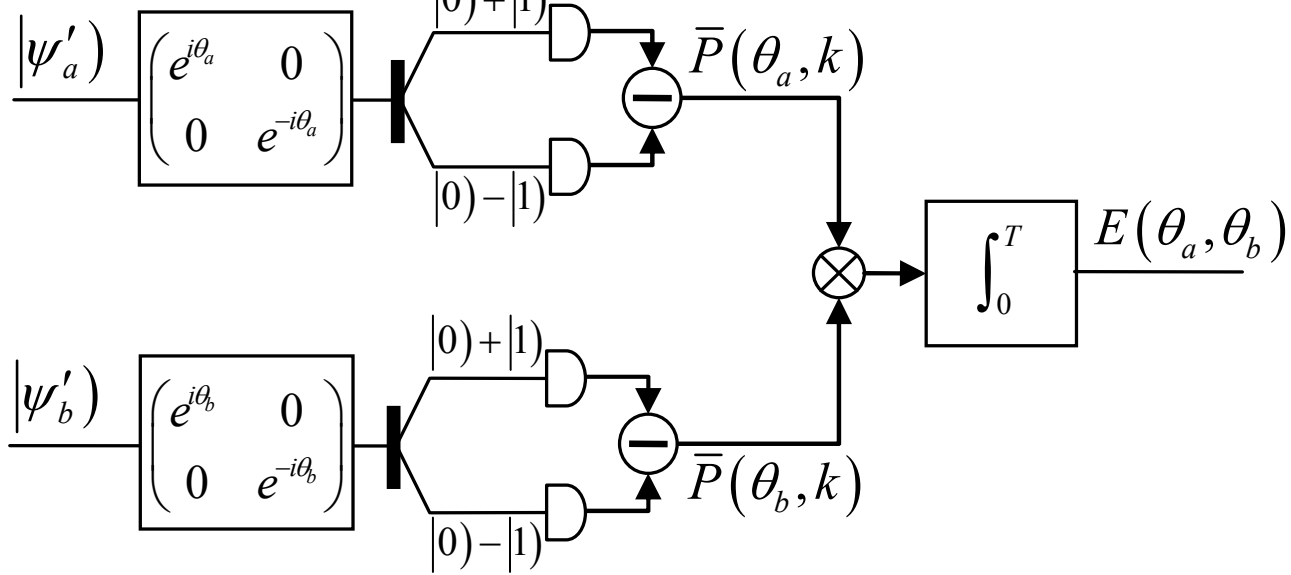
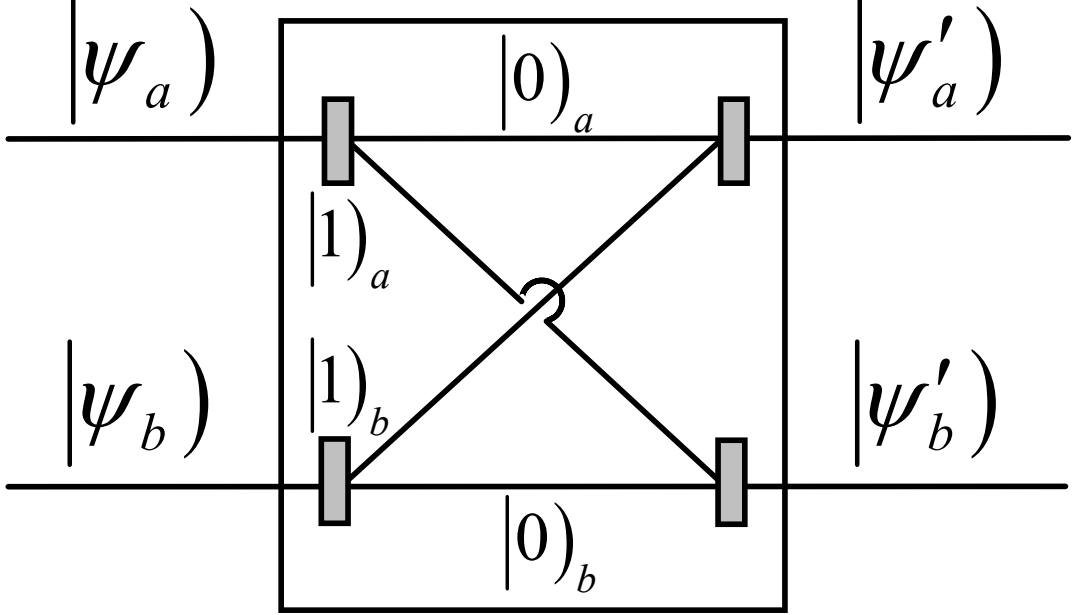
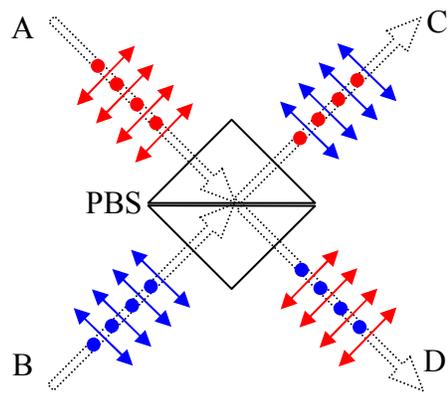
\end{document}